\documentclass[english,aps,prl,amsmath,amssymb,twocolumn,letterpaper,showpacs,superscriptaddress]{revtex4-1}
\usepackage{physics}
\usepackage{graphicx}
\usepackage{babel}
\usepackage{amstext}
\usepackage{esint}
\usepackage[unicode=true,pdfusetitle,
 bookmarks=true,bookmarksnumbered=false,bookmarksopen=false,
 breaklinks=false,pdfborder={0 0 1},backref=false,colorlinks=false]
 {hyperref}
\usepackage{pdfpages}
\usepackage{pgffor}

\makeatletter
\AtBeginDocument{\let\LS@rot\@undefined} 
\makeatother

\makeatother

\begin{document}

\title{Probing and dressing magnetic impurities in a superconductor}

\author{K. Akkaravarawong}
\affiliation{Department of Physics, University of California, Berkeley, CA 94720, USA}

\author{J. I. V\"ayrynen}
\affiliation{Microsoft Quantum, Microsoft Station Q, University of California, Santa Barbara, CA 93106-6105 USA}

\author{J. D. Sau}
\affiliation{Joint Quantum Institute, National Institute of Standards and Technology,
and University of Maryland, Gaithersburg, MD 20899, USA}

\author{E. A. Demler}
\affiliation{Department of Physics, Harvard University, Cambridge, MA 02138, USA}

\author{L. I. Glazman}
\affiliation{Department of Physics, Yale University, New Haven, CT 06520, USA}

\author{N. Y. Yao}
\affiliation{Department of Physics, University of California, Berkeley, CA 94720, USA}
\affiliation{Materials Science Division, Lawrence Berkeley National Laboratory, Berkeley, CA 94720, USA}

\date{\today}
\begin{abstract}
We propose a method to probe and control the interactions between an ensemble of magnetic impurities in a superconductor via microwave radiation. Our method relies upon the presence of sub-gap Yu-Shiba-Rusinov (YSR) states associated with the impurities.  Depending on the sign of the detuning, radiation generates either a ferro- or antiferromagnetic contribution to the exchange interaction. 
This contribution can bias the statistics of the random exchange constants stemming from the RKKY interaction. 
Moreover, by measuring the microwave response at the YSR resonance, one gains information about the  magnetic order of the impurities. 
To this end, we estimate the absorption coefficient  as well as the achievable strength of the microwave-induced YSR-interactions using off-resonant radiation. 
The ability to utilize microwave fields to both probe and control impurity spins in a superconducting host may open new paths to studying metallic spin glasses.
\end{abstract}
\date{\today}
\maketitle

The nature of interactions between magnetic impurities embedded in a metallic host gives rise to an intriguing state of matter: a spin glass~\cite{fischer_hertz_1991}. 
The RKKY exchange interaction between the impurities is carried by itinerant electrons and alternates in sign, depending on the inter-impurity separation ~\cite{PhysRev.96.99,10.1143/PTP.16.45,PhysRev.106.893}. 
The random position of the impurities with respect to each other results in a random-sign exchange interaction, frustrating the magnetic order in a system of localized spins. 
The efforts to understand the resulting low-temperature spin glass phase and the corresponding phase transition have led to the introduction of several important concepts in condensed matter physics, including the Edwards-Anderson~\cite{EA1975} and functional~\cite{Parisi1980} order parameters. 
Moreover, these efforts have also motivated a ever-expanding toolset of quantum control techniques aimed at directly controlling the interactions between magnetic impurities.

Remarkably, even the simplest spin glass model introduced by Sherrington and Kirkpatrick~\cite{Sherrington75} in direct analogy to the Curie-Weiss model of a ferromagnet turns out to be extremely rich and, unlike the Curie-Weiss model, not amenable to a straightforward mean-field theory treatment~\cite{Thouless-Anderson-Palmer,de-Almeida-Thouless}.  
The frustration of the magnetic moments manifests itself in both the thermodynamic and electron transport properties of a normal metal with a magnetic element dissolved in it. 
Starting with magnetic susceptibility measurements on AuFe alloys~\cite{PhysRevB.6.4220}, 
there are a substantial number of such studies performed on bulk samples \cite{doi:10.1063/1.1660392,doi:10.1063/1.2947017}.
With the development of mesoscopic systems, electron transport through mesoscale-sized alloys also received their fair share of attention; for example, the remanence of the resistance (i.e.~its dependence on the cooling protocol) of a mesoscopic AgMn device was investigated in~\cite{remanence}, while quantum interference effects in the conductance of CuMn and AgMn were studied respectively in \cite{PhysRevLett.66.2380}
and~\cite{PhysRevLett.111.187203}.

\begin{figure}
	\centering
	\begin{tabular}{cc}
		\includegraphics[width=0.35\columnwidth]{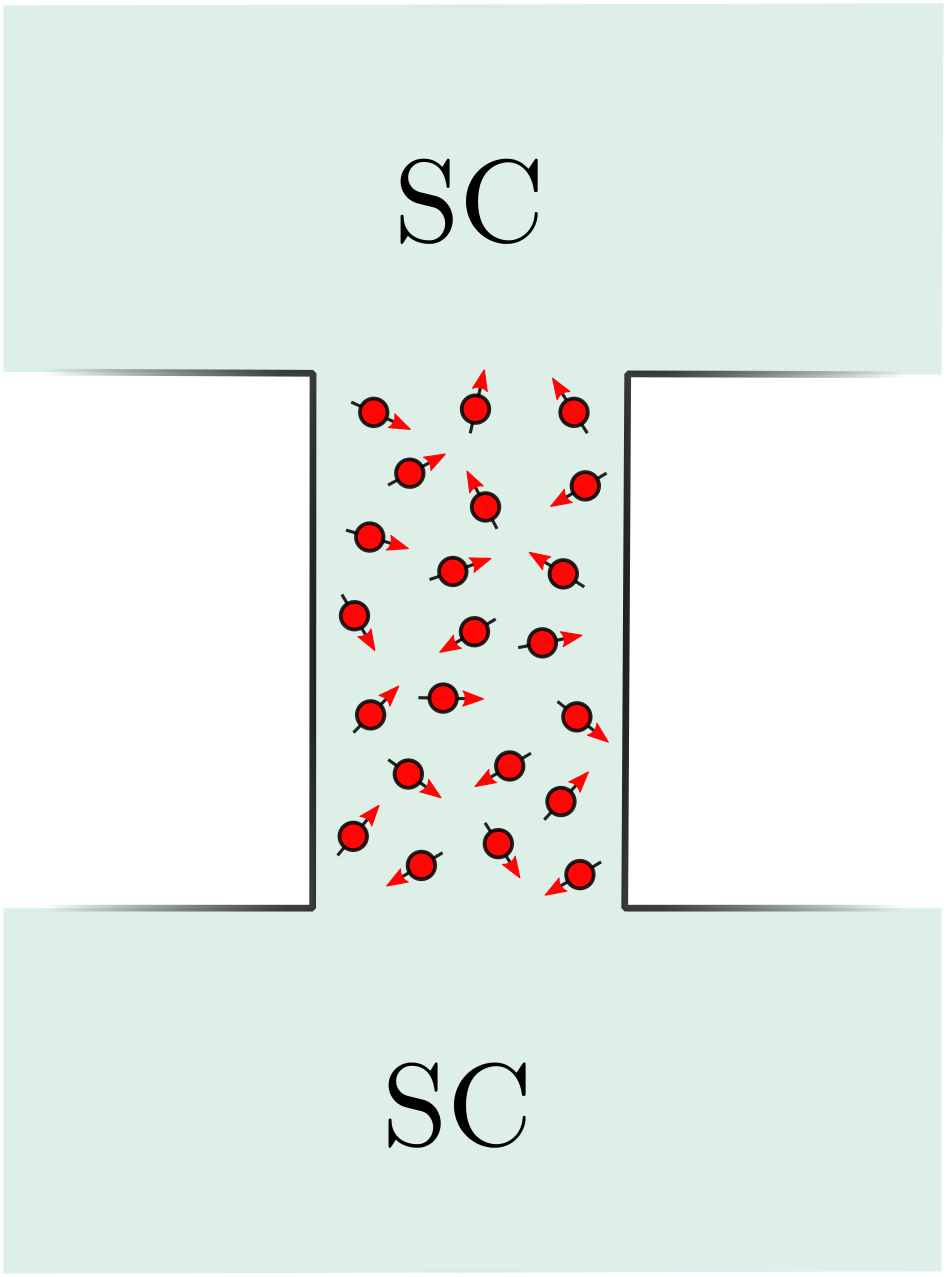}& \quad 
		\includegraphics[width=0.55\columnwidth]{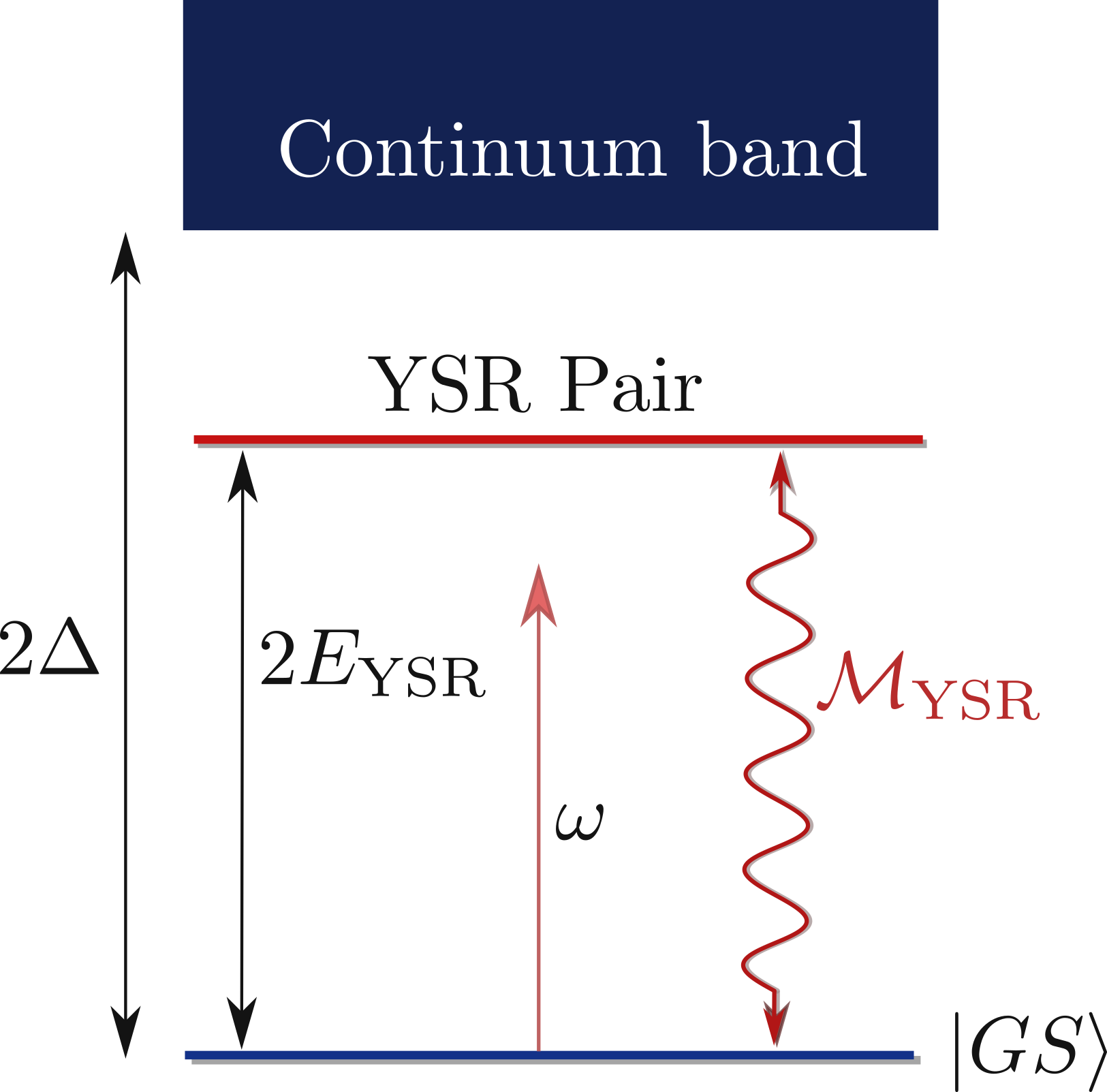}
	\end{tabular}
	\caption{(a) Schematic depicting an S-s-S junction consisting of two large superconducting banks connected by a narrow constriction, with the arrows representing magnetic impurities.
		(b) Shows the energy levels of the YSR states associated with two magnetic impurities in a superconductor. The radiation matrix element $\mathcal{M}_{\textrm{YSR}}$ depends on the mutual spin orientation of the impurities (i.e.~vanishes for parallel magnetic moments), resulting in a spin-dependent AC Stark shift that translates to an effective, microwave-induced, spin-spin interaction. 
	} \label{fig:levels}
\end{figure}

In this Letter, we explore a novel method for investigating mesoscopic spin glasses using techniques recently perfected  in the development of superconducting qubit technologies~\cite{Devoret1169}. 
In particular, we consider the possibility of utilizing microwave radiation to directly probe and possibly control the many-body state of an ensemble of magnetic moments embedded in a thin superconducting bridge (Fig.~\ref{fig:levels}). 

Much as in a normal metal, magnetic impurities in a superconductor also exhibit random-sign RKKY interactions. 
This RKKY interaction is hardly modified by superconductivity, so long as the typical distance, $d$, between the  impurities is shorter than the superconducting coherence length, $\xi$.
For impurities separated by larger distances, the interaction is instead antiferromagnetic and dominated by a virtual process involving YSR states~\cite{yaoPRL14}; however, at such distances, the interactions are typically weak since they decay exponentially with $d/\xi$.
Herein lies intuition behind our approach: To utilize microwave driving to enhance the virtual hybridization between the superconducting condensate and the YSR states.

With respect to such microwave excitation, there are two main differences between normal-metal and superconducting hosts. The first is that there exists a gap, $\Delta$, in the spectrum of excitations in a superconductor. In the absence of a magnetic field and impurities, a conventional s-wave superconductor, such as aluminum, possesses time-reversal symmetry. As a result, the gap is ``hard": At low temperatures there is a frequency threshold, $\omega_{\rm th}=2\Delta/\hbar$, for the absorption of electromagnetic radiation. The second difference is that a magnetic impurity in a superconductor creates a localized Yu-Shiba-Rusinov (YSR) state with energy $E_{\rm YSR}$ within the gap~\cite{YU65,Shiba1968,Rusinov1969,PhysRevLett.117.186801,HEINRICH20181,PhysRevLett.115.087001,PhysRevLett.120.156803,PhysRevLett.118.117001,PhysRevLett.121.196803}. 
A single YSR state may host no more than one quasiparticle, and therefore cannot facilitate absorption from the condensate. However, a pair of YSR states separated by distances $\lesssim\xi_{\text{YSR}}$ creates a discrete-energy state for an electron pair where $\xi_{\rm YSR}=\xi\sqrt{\Delta/(\Delta-E_{\rm YSR})}$ is the characteristic length-scale of a YSR state. 
To this end, at low temperatures, the sub-gap absorption results from a process in which a microwave photon transfers a Cooper pair from the condensate onto the pair of YSR states, leading to an absorption line centered at $\omega=2E_{\rm YSR}/\hbar$. 

Crucially, the magnitude of this absorption by a YSR pair depends on the mutual orientation of the magnetic moments. 
For moments oriented in parallel, the associated pair of YSR states \emph{cannot} accept a singlet Cooper pair (for simplicity, we assume that there is no spin-orbit coupling). 
Thus, the absorption is maximized for antiparallel moments and varies as $F[\mathbf{S}(\mathbf{R}_1), \mathbf{S}(\mathbf{R}_2)] =1-\hat{\textbf{S}}_1\cdot \hat{\textbf{S}}_2 $, where $\hat{\textbf{S}}_{1,2} = \textbf{S}_{1,2}/S$ are the unit vectors indicating the orientation of the magnetic moments (treated classically).
Therefore, the sub-gap absorption coefficient provides information regarding ferromagnetic order at the scale, $|\mathbf{R}_1- \mathbf{R}_2| \lesssim \xi_{\text{YSR}}$. 
The absorption line width and its detailed shape depend on the inevitable spread of the contact exchange interaction ~\cite{Yazdani1767,2008PhRvL.100v6801J,PhysRevLett.119.197002,PhysRevLett.119.197002,2018arXiv181111591K} and the overlap between the YSR states~\cite{PhysRevB.84.224517,Kezilebieke2018,PhysRevLett.120.156803,PhysRevLett.120.167001}.

For an ensemble of moments with density $n \gtrsim \xi^{-3}$, the many-body ferromagnetic order can be deduced from $\bar{F}(\{\mathbf{S}(\mathbf{r})\}) =\langle \sum_{i \neq j} K(|\mathbf{R}_i-\mathbf{R}_j|)F[\mathbf{S}(\mathbf{R}_i), \mathbf{S}(\mathbf{R}_j)] \rangle/N$ where $\langle \dots \rangle$ denotes averaging over YSR state energies, $K(r) \sim \Theta(\xi_{\text{YSR}}-r)$ [$\Theta$ is the Heaviside step function], $N \sim n^2 \xi^3 \mathcal{V} (w/\xi)^{3-D}$ is a normalization factor and $\mathcal{V}$ is the volume of the sample which we think of as either a wire ($D=1$) or as a film ($D=2$) with transverse dimension $w \ll \xi$. This quantity is related to the dissipative part of the conductivity integrated over the absorption line:
\begin{equation}
\int_0^{2\Delta/\hbar} d\omega \sigma(\omega)= \bar{F}(\{{\bf S}({\bf r})\})\Sigma\,,\quad \Sigma\sim\sigma_n\frac{n^2}{\nu^2\Delta} \frac{1/n \xi^2}{l+ 1/n \xi^2}\,,
\label{sigma}
\end{equation}
where $n$ is the impurity concentration, $\sigma_n$ and $\nu$ are, respectively, the normal-state conductivity and the density of states at the Fermi-level, and $l$ is the electron elastic mean free path; note that the last factor in Eq.~(\ref{sigma})  extrapolates between the regimes of long and short mean free paths.   

\begin{figure}
\centering
\includegraphics[width=3in]{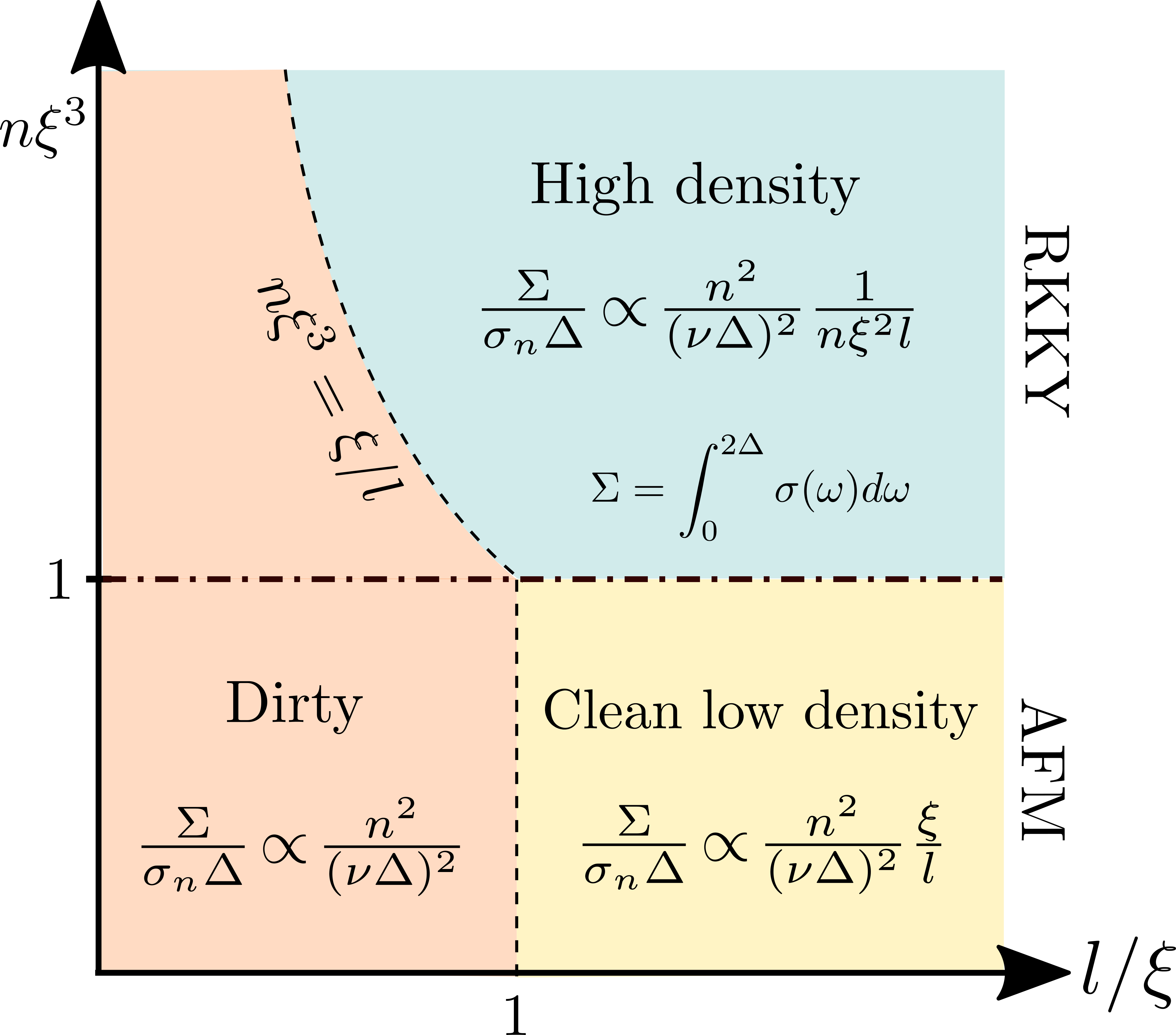}
\caption{Schematic cross-over diagram of the integrated sub-gap conductivity $\Sigma$ 
as a function of the elastic mean free path $l$ and the density of magnetic impurities $n$. 
	In the clean and dense  limit (upper right), the YSR states are strongly hybridized and $\Sigma \propto n$~\cite{PhysRevB.84.224517}.
	At lower densities ($n \xi^3 \gg 1$), the hybridization is negligible and the absorption requires two  magnetic impurities within $\xi$  of each other, $\Sigma \propto n^2$ (bottom right). 
	In the dirty case, $l \ll \min(\xi, 1/n \xi^2)$, the dependence of $\Sigma$ on $n$  is similarly quadratic (left).
	In the dense regime, $n\xi^3 \gg 1$ (top), the interactions between the magnetic impurities are dominated by RKKY mechanism (favoring a spin glass phase), while at low density, $n\xi^3 \ll 1$ (bottom), the exchange interaction is exponentially weak and antiferromagnetic. 
}  \label{fig:diagram}
\end{figure}

As aforementioned, in the absence of microwaves, there are two components of the inter-impurity interaction carried by virtual excitations of the itinerant electrons. The first (and dominant at $d\lesssim\xi$) one comes from the continuum of Bogoliubov quasiparticles and is responsible for the conventional indirect exchange coupling, {\it i.e.}, RKKY interaction in normal metals and its counterpart in superconductors. The second one owes to the discrete YSR states and is specific to superconductors. 
Borrowing the idea of ``off-resonant dressing" from quantum optics~\cite{cohen1992atom}, we note that off-resonant microwave radiation creates an \emph{additional} channel for virtual transitions of Cooper pairs onto a pair of YSR states. 
For sufficiently strong drives, the corresponding amplitude may successfully compete with the one existing in the absence of radiation \cite{yaoPRL14}, while the conventional RKKY component is only weakly affected, so long as the radiation remains far detuned from the gap edge. 
The sign of the effective interaction induced by the off-resonant drive depends on the sign of the detuning, $\omega-2E_{\rm YSR}$, and is ferromagnetic for $\omega-2E_{\rm YSR}>0$. 

One can estimate the strength of the microwave-induced interaction, $J_{\rm YSR}^{\rm ind}$, relative to the RRKY component $J_{\rm RKKY}$ as
\begin{equation}
\frac{\langle J_{\rm YSR}^{\rm ind} \rangle}{\langle J_{\rm RKKY} \rangle }\sim\frac{\Delta}{2E_{\rm YSR}-\omega}\left(\frac{d}{\xi}\right)^3\left(\frac{e \mathcal{E}\xi}{\omega}\right)^2\,,
\label{JYSR}
\end{equation}
where $\mathcal{E}$ is the electric field of the microwave. In order for superconductivity to remain intact, the last factor here must be small; in order to avoid resonant absorption, the denominator of the first factor must exceed the YSR absorption line width. Together, these two conditions set a limit for the strength of the ``dressed'' interaction. Intriguingly, in a dilute system, at $d\sim\xi$, the ``dressed'' interaction may compete with the conventional RKKY component opening up the possibility of studying the de Almeida -- Thouless line~\cite{de-Almeida-Thouless} in a spin glass.

{\emph{Model}}---Our starting point is the BCS Hamiltonian of an s-wave superconductor with magnetic impurities. The Bougoliubov-de Gennes (BdG) Hamiltonian takes the form
\begin{equation}
\mathcal{H} = \epsilon_{p} \tau_{z} + \Delta \tau_{x} - \sum_{i} J_i  \textbf{S}_{i} \cdot \pmb{\sigma} \delta (\textbf{r} - \textbf{R}_{i})\,, \label{eq:H0}
\end{equation} 
where $\epsilon_{p} =  -\frac{1}{2m} \nabla^2 - \mu$ is the kinetic energy (we set $\hbar = 1$), $\mu$ the chemical potential, and $\Delta$ the superconducting order parameter. The Hamiltonian $H = \int d\textbf{r} \Psi^{\dagger} \mathcal{H} \Psi/2$ 
 is written in conventional Nambu spinor notation, where $\Psi = [\psi_{\uparrow},\psi_{\downarrow},\psi_{\downarrow}^{\dagger},-\psi_{\uparrow}^{\dagger}]^{T}$ and  $\tau(\sigma)$ are Pauli matrices acting on the particle-hole (spin) space. The last term in the Hamiltonian represents the contact interaction between electrons and the impurity spins, where $J_i$ characterizes the coupling strength and $\textbf{S}_{i}$ is the spin of the \textit{i}$^{\textrm{th}}$ impurity located at  position $\textbf{R}_i$.

The energy of the sub-gap YSR bound state localized around  impurity $i$ is $E^{i}_{\text{YSR}} = \Delta{(1-\alpha_i^2)}/{(1+\alpha_i^2)}$~\cite{YU65,Shiba1968,Rusinov1969}, where $\alpha_i = \nu \pi J_i S/2$ is a dimensionless exchange coupling. 
We treat the impurity spins classically assuming they have equal magnitudes $S$ but in general different orientations. 
The 4-component eigenspinors of the BdG Hamiltonian $\mathcal{H}$ corresponding to the  sub-gap YSR states are given by

\begin{equation}
\Phi^+_{i,\textbf{S}_i }(\textbf{r}) = \frac{1}{\mathcal{N}_i} \frac{e^{- \abs{\sin{2\delta_i}}r_i/\xi}}{k_{F}r_i} U(\hat{\textbf{S}}_i) \begin{pmatrix} \sin{(k_{F}r_i + \delta_i)} \ket{\uparrow} \\ \sin{(k_{F}r_i - \delta_i)} \ket{\uparrow} \end{pmatrix}\,,
\end{equation}
and $\Phi^-_{i,\textbf{S}_i} = CT\Phi^+_{i,\textbf{S}_i }$
where a particle ${(+)}$ and hole ${(-)}$  states $\Phi_{i,\textbf{S}_i}^{\pm}$ have energies $\pm E^i_{\text{YSR}}$ and are related by the antiunitary symmetry transformation
 $CT = \tau_y \sigma_y K$ where  $K$ denotes  complex conjugation; $\mathcal{N}_i^2= 2\pi\nu \Delta \alpha_i/ (1+\alpha_i^2)$ is a normalization factor, $r_i = {|\textbf{r} - \textbf{R}_i|}$ the relative distance from the impurity,   $\delta_i = \tan^{-1}(\alpha_i)$ the phase shift, $\ket{\uparrow}$ is the $+1$ eigenstate of $\sigma_z$ and $U(\hat{\textbf{S}}_i)$ is a unitary rotation operator aligning the quantization axis of the Nambu spinor $\Psi$ with the direction of the impurity spin.
As we are interested in the physics resulting from low energy microwave excitation, we will project the electron field operator onto these sub-gap YSR states,
\begin{equation}
\Psi(\textbf{r})
= \sum_{i} \left[ \Phi_{i,\textbf{S}_i}^+(\textbf{r})\gamma_{i} + \Phi_{i,\textbf{S}_i}^-(\textbf{r})\gamma_{i}^{\dagger} \right]\,, \label{eq:expand}
\end{equation}
where $\gamma_i$ is the annihilation operator of a YSR state located at the \textit{i}$^{\textrm{th}}$ impurity. The projected Hamiltonian then becomes $H = \sum_i E_i \gamma^{\dagger}_i\gamma_i +\sum_{i,j}( \mathcal{M}_{ij}\gamma_i^{\dagger} \gamma_j^{\dagger} +\mathcal{M}'_{ij}\gamma_i^{\dagger} \gamma_j+ h.c.)$. To be specific, we focus  on the low-density regime, $n\xi_{\text{YSR}}^3 \ll 1$, which exhibits the same qualitative behavior as the high-density regime. 
In the low-density regime, we can neglect 
the hopping 
and  non-radiative pair-creation terms from $H$ which simplifies the calculation significantly. 

{\emph{Radiation matrix element}}---We now turn to calculating the matrix element, $\mathcal{M}_{\text{YSR}}$, corresponding to radiation-assisted YSR-pair-creation. When the system is coupled to a  weak microwave field, the vector potential $\tilde{\textbf{A}}$ enters as $\frac{\hbar} {i} \nabla \rightarrow \frac{\hbar}{i}\nabla +\frac{e}{c} \tilde{\textbf{A}}$ and the superconducting order parameter is  generally both complex and spatially dependent, i.e. $\Delta(\textbf{r}) =|\Delta| e^{i \theta(\textbf{r})}$. 
We choose to work in the London gauge where the order parameter is real and the integral of the new vector potential, $\textbf{A} = \tilde{\textbf{A}} - \frac{\hbar c}{e}\nabla \theta(\textbf{r})$, yields a gauge-invariant phase difference that gives rise to the supercurrent in a superconductor.

The electromagnetic perturbation to the BdG Hamiltonian is given by: $H_{\text{EM}} = \int d^3 \textbf{r} \, (e/2)\Psi^{\dagger}(\textbf{A} \cdot \textbf{v} + \textbf{v} \cdot \textbf{A})\Psi$, where $\textbf{v}$ is the velocity operator. Using Eq.~(\ref{eq:expand}), the projected perturbation Hamiltonian takes form $H_{\text{EM}} \approx \mathcal{M}_{\text{YSR}}^{1,2} \gamma_1^{\dagger} \gamma_2^{\dagger} + h.c.$ where $\mathcal{M}_{\text{YSR}}^{1,2} = \int d^3 \textbf{r} \, \textbf{J}_{1,2} \cdot \textbf{A} $.
We assume a  negligible thermal population of YSR states and thereby ignore  hopping terms $\gamma_1^{\dagger} \gamma_2$. 
The current density $\textbf{J}_{1,2}$ is given by
\begin{equation}
\textbf{J}_{1,2}(\textbf{r})\!
= \!\frac{e \hbar}{2mi} \! \left\{ \Phi_{1,\textbf{S}_1}^{+ }(\textbf{r}) \nabla \Phi^-_{2,\textbf{S}_2}(\textbf{r})\! - \Phi_{2,\textbf{S}_2}^{+}(\textbf{r}) \nabla \Phi^-_{1,\textbf{S}_1}(\textbf{r}) \right\}. \label{eq:current}
\end{equation}
Since the relevant microwave frequencies ($\omega \leq 2 \Delta$) corresponds to wavelengths significantly longer than both the superconducting coherence length and the characteristic YSR length-scale, $\textbf{A}$ can be treated as position-independent. 
Moreover, since the integration domain contains all of space, the integral of the anti-symmetric portion of the integrand vanishes.
The integrand can therefore be symmetrized, 
\begin{equation}
\mathcal{M}_{\text{YSR}}^{1,2} = \frac{1}{2} (\textbf{A} \cdot \hat{\textbf{R}}) \, \hat{\textbf{R}} \cdot  \! \int d^3 \textbf{r}   \big[ \textbf{J}_{1,2}(\textbf{r}) + \textbf{J}_{1,2}(-\textbf{r})  \big] \,,\label{eq:Mintegral}
\end{equation}
where we have also used the fact that the current density, Eq.~(\ref{eq:current}), is rotationally symmetric around the axis connecting the two impurities, $\textbf{R} = \textbf{R}_2 - \textbf{R}_1$ and $\hat{\textbf{R}} = \textbf{R}/|\textbf{R}|$, and therefore only the component of the vector potential that is parallel to this axis contributes to absorption. 

Owing to the rotational symmetry, the integral is effectively two-dimensional and can be done in elliptical coordinates~\cite{Note1}, 
$r_{+} = (r_1 +r_2)/2$,   $ r_{-} = r_1 - r_2 $, 
where the integral measure becomes
$\int d^3 \textbf{r} \rightarrow \pi \int_{R/2}^{\infty} dr_+\int_{-R}^{R}dr_- \{ r_+^2 - \frac{r_-^2}{4}  \}$.
In these new coordinates, $r_+^2 - r_-^2/4  = r_1 r_2$, exactly canceling out the power-law decay of the YSR wave function!

For two \textit{identical} impurities, the current density respects an additional reflection symmetry about the plane perpendicular to and bisecting $\textbf{R}$. 
This symmetry would imply that the integral in Eq.~(\ref{eq:Mintegral}) vanishes. 
Thus, a non-zero radiation matrix element requires the breaking of this reflection symmetry. 
In practice \cite{PhysRevLett.119.197002}, this is always the case as  one invariably  observes fluctuations in the exchange coupling strengths, suggesting that the reflection symmetry is naturally broken by disorder effects. 
Therefore, we assume hereafter that the impurities are not identical, $\alpha_{1}\neq \alpha_{2}$.

After a straightforward but tedious calculation, the general expression of the matrix element can be obtained.
Assuming that $| \, |\sin 2\delta_1|-|\sin 2\delta_2| \, | \ll \xi/R$, the matrix element can be expanded to first order in the coupling difference, $|\alpha_1 -\alpha_2|$, and $1/(k_F R)$ as~\footnote{See Supplementary Material for a derivation of the pair transition matrix element, the integrated conductivity in the clean dilute limit as well as  calculation of the ratio $J_{\text{YSR}}^{\text{ind}}  / J_{\text{RKKY}}  $.}   
\begin{eqnarray}
|\mathcal{M}_{\text{YSR}}^{1,2}| \approx&   \Delta ~ \sqrt[]{ F(\textbf{S}_1,\textbf{S}_2)}   \, \dfrac{|e \vec{\mathcal{E}} \cdot \hat{{\textbf{R}}}| \xi}{ h \omega} \dfrac{|\sin{k_F R}|}{k_F \xi}   \nonumber \\
& \quad \times e^{-R/\xi_{\text{YSR}}} |\alpha_1 -\alpha_2| \dfrac{2\sqrt{2} \alpha^3}{(1+\alpha^2)^3}\,, \label{eq:matrix element}
\end{eqnarray}
where $\vec{\mathcal{E}} = -\partial \textbf{A} /\partial t$ is the electric field of the applied microwaves  and $\alpha = (\alpha_1 +\alpha_2)/2$, the average coupling strength of the two impurities. 
One important and intriguing observation: Due to the longer intrinsic YSR length scale  $\xi_{\text{YSR}} = \xi/|\sin2\delta|$ as well as the absence of a power-law decay in the matrix element, the corresponding microwave-induced  interaction has a \emph{significantly} longer range than both the  RKKY interaction and the bare YSR interaction~\cite{yaoPRL14} (in the absence of microwaves).

{\emph{Experimental Implementation}}---In this section, we propose an experimental implementation based upon superconducting circuits, which enables one to utilize microwave fields to both probe the spin state of the impurities as well as control their effective  interactions. 
Since our proposed  microwave-dressed interactions require both a superconducting background and a supercurrent, a natural setup is an S-s-S junction created from two large superconducting leads linked by a constriction (Fig.~\ref{fig:levels}); such a setup has previously been used in experiments probing Andreev bound states~\cite{PhysRevLett.112.047002}.
 An oscillating bias potential, $V \cos {\omega t}$, can then be applied to the leads, to create a time-dependent supercurrent $j \propto \frac{\hbar e }{m} \nabla \theta$ governed by the Josephson relation $\partial_t \theta = \frac{2eV}{\hbar} \sin {\omega t}$, where $\theta$ is the gauge-invariant phase difference between the leads and $\omega$ is the microwave frequency.

To probe the many-body state of the impurities, we propose to apply a \emph{resonant} microwave drive so that the ordering can be inferred from the integrated dissipative part of the sub-gap conductivity, $\int_0^{2\Delta} \sigma(\omega) d\omega $. The dissipative conductivity $\sigma$ is related to the energy absorption rate $\omega \Gamma =  \sigma  \mathcal{E}^2/2$, where $\Gamma$ is the transition rate obtained by plugging Eq.~(\ref{eq:matrix element}) into  Fermi's golden rule. 
At low temperatures, the initial state consists of unoccupied YSR states. The integrated sub-gap conductivity depends on the average distance between impurities and the elastic mean free path $l$. It can be analyzed in various limits that are detailed below and summarized in Fig.~\ref{fig:diagram}.

In the low-density limit ${n \xi_{\text{YSR}}^{3} \lesssim 1} $, the YSR states are well-localized so that the hybridization-induced energy splitting of the YSR states is negligible compared to $\Delta$. The  sub-gap conductivity can be written as a sum of pair-creating transitions, 
\begin{equation}
\sigma(\omega) = \frac{2}{\mathcal{V}\mathcal{E}^2}\sum_{i > j} 2\pi \omega | \mathcal{M}_{\text{YSR}}^{i,j}|^2 \delta(\omega - E^i_{\text{YSR}}-E^j_{\text{YSR}}) \,, \label{eq:conductivity}
\end{equation}
where $\mathcal{M}_{\text{YSR}}^{i,j}$ is the radiation-assisted matrix element to create a  pair of YSR quasiparticles on the \textit{i}$^{\textrm{th}}$ and \textit{j}$^{\textrm{th}}$ impurity. Assuming a uniform distribution of uncorrelated YSR levels in a narrow band, $W \ll 2E_{\text{YSR}}$, and ensemble-averaging the conductivity, we find
~\cite{Note1}
\begin{align}
\bigg\langle \frac{\sigma(\omega)}{\sigma_{n}} \bigg\rangle   
= &\frac{\bar{F}(\{ {\textbf{S}(\textbf{r}}) \})}{\hbar\omega}\,\frac{2 \alpha^4}{3 (1+\alpha^2)^2} \frac{n^2}{\hbar \pi \nu^2 \Delta} (w/\xi)^{3-d}
\nonumber \\
 &\quad \times \frac{\xi }{l}  \frac{1}{2W \Delta} g(\hbar \omega - 2\bar{E}_{\text{YSR}},W) \, 
, \label{eq:sigmaexact}
\end{align}
where $\bar{\alpha}$ is the mean value of $\alpha_i$ and $ \bar{E}_{\text{YSR}}$ the average YSR energy. 
The normalized distribution function $g(\omega,W) = 2(1 -|\omega|/W)^3 \,  \Theta(W - |\omega|)/W$ (specific to a uniform YSR band) characterizes the energy-dependence of the absorption and has a peak of width $W$.
We normalize the conductivity by its normal-state value $\sigma_n = 2e^2\nu\frac{1}{2}v_F l$, where $l$ is the electron mean free path from non-magnetic impurities; 
Eq.~(\ref{eq:sigmaexact}) is valid in the  clean limit $l \gg \xi$. 
The dirty limit $l \ll \xi$ is obtained by replacing $\xi \to l$ in the second line of Eq.~(\ref{eq:sigmaexact})~\cite{PhysRevB.84.224517,PhysRevB.96.134501}.

In the high-density regime, $n\xi_{\text{YSR}}^3 \gg 1$, the YSR states will strongly hybridize. 
The conductivity in this case is derived in~\cite{PhysRevB.84.224517}. 
At a qualitative level, the high-density limit can be obtained from  Eq.~(\ref{eq:sigmaexact}) by replacing $\xi$ (in the second line)  by an effective mean free path arising from magnetic impurities, $\xi \to 1/(n \xi^2)$. 
When the above conductivity is integrated over the sub-gap states, one naturally recovers Eq.~(\ref{sigma}).

Finally, we now estimate the achievable strength of the dressed interactions induced by an off-resonant microwave field. The relevant energy levels and matrix element are depicted  in Fig.~\ref{fig:levels}b. At leading order, one finds that the radiation-assisted YSR pair creation results in an effective spin-spin interaction originating from a spin-dependent AC stark shift to the ground state energy,
\begin{equation}
H_{\text{eff}} = \delta E_{\text{GS}} = \frac{|\mathcal{M}_{\text{YSR}}|^2}{2  E_{\text{YSR}}- \omega}=  J_{\text{YSR}}^{\text{ind}} ~ \textbf{S}_1 \cdot \textbf{S}_2,
\end{equation}	
where we have neglected a spin-independent overall shift.  
To compare the relative strength of this dressed interaction with the RKKY component, we use Eq.~(\ref{eq:matrix element}) and the expression for the RKKY interaction from Ref.~\cite{yaoPRL14}, resulting~\cite{Note1} in the estimate presented in Eq.~(\ref{JYSR}).

In summary, we have shown that the electronic sub-gap states hosted by magnetic impurities in a superconductor provide a new way to access  magnetic order.
This opens the possibility to probe and control metallic spin glass physics in a superconducting narrow-bridge junction  by using microwave driving (Fig.~\ref{fig:levels}a). 
Keeping the transverse size $w$ of the bridge thinner than the London length ensures that the supercurrent  is approximately uniform and couples to the magnetic moments in the full volume. 
Unlike conventional Andreev bound states, the YSR levels are insensitive to a static phase difference across the junction which provides a simple way to distinguish the respective contributions to the dissipative conductivity. 
Looking forward, our work also opens the door to an intriguing quantum information platform where magnetic impurities in a superconducting host play the role of quantum memories, while microwave driving can lead to on-demand long-range gates \cite{PhysRevLett.89.147902}. Here, the absence of a power-law decay in the radiation matrix element could enable all-to-all connectivity between qubits as well as multi-body interactions, both of which are important for reducing the gate depth of certain quantum algorithms \cite{BernsteinVazirani, HiddenShift}.

\emph{Acknowledgments}---We gratefully acknowledge the insights of and discussions with K. Murch, I. Siddiqi, and T. Ojanen. 
The authors would particularly like to thank G. Zarand for previous collaborations on related work and for pointing out the impact of reflection symmetry.
 This work was supported by the U.S. Department of Energy, Office of Basic Energy Sciences, Division of Materials Sciences and Engineering under award AWD00003522, the  David and Lucille Packard Foundation and the Sloan Foundation.
JS acknowledges support from the NSF-DMR1555135 (CAREER).
ED acknowledges support from the Harvard-MIT CUA, the AFOSR-MURI  Photonic Quantum Matter (award FA95501610323), and the DARPA DRINQS program (award D18AC00014).
LG acknowledges support from the NSF DMR Grant No.~1603243.

\bibliographystyle{apsrev4-1}
\bibliography{refs}

\end{document}


\title{Supplementary material for ``Probing and dressing magnetic impurities in a superconductor''}

\author{K. Akkaravarawong}
\affiliation{Department of Physics, University of California, Berkeley, California 94720, USA}

\author{J. I. V\"ayrynen}
\affiliation{Microsoft Quantum, Microsoft Station Q, University of California, Santa Barbara, California 93106-6105 USA}

\author{J. D. Sau}
\affiliation{Joint Quantum Institute, National Institute of Standards and Technology,
and University of Maryland, Gaithersburg, MD 20899, USA}

\author{E. A. Demler}
\affiliation{Department of Physics, Harvard University, Cambridge, MA 02138, USA}

\author{L. I. Glazman}
\affiliation{Department of Physics, Yale University, New Haven, Connecticut 06520, USA}

\author{N. Y. Yao}
\affiliation{Department of Physics, University of California, Berkeley, California 94720, USA}
\affiliation{Materials Science Division, Lawrence Berkeley National Laboratory, Berkeley, CA 94720, USA}

\date{\today}

\maketitle

In this supplementary material we show in detail the derivation of the pair transition matrix element, the integrated conductivity in the clean dilute limit as well as the calculation of the ratio $J_{\text{YSR}}^{\text{ind}}  / J_{\text{RKKY}}$. In addition, we compare our analytic result to direct numerical integration.

\section{Calculation for the matrix elements}

\subsection{Symmetry argument for vanishing perpendicular field contribution}
From Eq.~(6) of the main text, the current density is rotationally symmetric around $\hat{\textbf{R}}$:
\begin{align}
\textbf{J}(\textbf{r}) = \mathcal{O}\textbf{J}(\mathcal{O}\textbf{r})
\end{align}
where $\mathcal{O}$ is the rotation matrix about the axis  $\hat{\textbf{R}}$. We can decompose the vector potential into the parallel and perpendicular components: $\textbf{A}_{\parallel}= (\textbf{A} \cdot \hat{\textbf{R}})\hat{\textbf{R}}$ and $\textbf{A}_{\perp} = \textbf{A}-\textbf{A}_{\parallel}$ respectively.
Due to the rotational symmetry, the integral of the perpendicular vector potential vanishes.
\begin{align}
\mathcal{M}_{\text{YSR}}^{\perp}
&= \int d^3 \textbf{r} \textbf{A}_{\perp} \cdot \textbf{J}(\textbf{r}) = \int d^3 \textbf{r}  \textbf{A}_{\perp} \cdot \tilde{\mathcal{O}} \textbf{J}(\tilde{\mathcal{O}} \textbf{r}) 
= \int d^3 \textbf{r}' ( \tilde{\mathcal{O}}^{-1}\textbf{A}_{\perp}) \cdot \textbf{J}(\textbf{r}') 
= \int d^3 \textbf{r}' ( -\textbf{A}_{\perp}) \cdot \textbf{J}(\textbf{r}')
= 0\,,
\end{align}
where $\textbf{r}' = \tilde{\mathcal{O}}\textbf{r}$ and we choose $\tilde{\mathcal{O}}$ such that $\tilde{\mathcal{O}}^{-1}\textbf{A}_{\perp} = -\textbf{A}_{\perp}$ which is possible for any vector perpendicular to $\hat{\textbf{R}}$.

\subsection{Setting up the matrix element integral}

By using Eqs.(6)-(7) of the main text, and defining the directional derivative $\hat{\textbf{R}} \cdot \nabla = \nabla_{\textbf{R}}$, the matrix element can be written as $\mathcal{M}_{\text{YSR}}^{1,2}
= -\frac{e\hbar}{4mi}(\textbf{A} \cdot \hat{\textbf{R}})I(k_F R, \frac{R}{\xi})$ where the integral is  
\begin{align}
I(k_F R, \frac{R}{\xi}) =
\int d^3\textbf{r} \biggl( 
&\Phi_{1,\textbf{S}_1}^{+\dagger}(\textbf{r}) \nabla_{\textbf{R}}  \Phi_{2,\textbf{S}_2}^{-}(\textbf{r})
+ \nabla_{\textbf{R}} \Phi_{2,\textbf{S}_2}^{+\dagger}(\textbf{r})  \Phi_{1,\textbf{S}_1}^{-}(\textbf{r}) \nonumber \\
& \quad -  \nabla_{\textbf{R}} \Phi_{1,\textbf{S}_1}^{+\dagger}(-\textbf{r}) \Phi_{2,\textbf{S}_2}^{-}(-\textbf{r}) 
- \Phi_{2,\textbf{S}_2}^{+\dagger}(-\textbf{r})  \nabla_{\textbf{R}} \Phi_{1,\textbf{S}_1}^{-}(-\textbf{r}) \biggr)\,.
\end{align}
Here we integrated-by-parts in the second and third term in the parentheses which resulted in  minus signs (the boundary terms vanish because the eigenspinors are localized). 
To greatly simplify the calculation, we take the derivative outside the integral by the following procedures: first, shift the coordinate origin of the eigenspinors in the first and the second line of Eq.~(S3) to be at impurity 1 and 2 respectively. 
(Note that Eq.~(4) of the main text defines the eigenspinor as an implicit function of relative distance, $r_i = |\textbf{r}- \textbf{R}_i|$.)
The integral becomes
\begin{align}
I(k_F R, \frac{R}{\xi})
= \int d^3 \textbf{r}
\biggl(  &\Phi_{1,\textbf{S}_1}^{+\dagger}(|\textbf{r}|) \nabla_{\textbf{R}}  \Phi_{2,\textbf{S}_2}^{-}(|\textbf{r}-\textbf{R}|) 
+ \nabla_{\textbf{R}} \Phi_{2,\textbf{S}_2}^{+\dagger}(|\textbf{r}-\textbf{R}|)  \Phi_{1,\textbf{S}_1}^{-}(|\textbf{r}|)  \nonumber \\
&\quad - \nabla_{\textbf{R}} \Phi_{1,\textbf{S}_1}^{+\dagger}(|\textbf{r}-\textbf{R}|) \Phi_{2,\textbf{S}_2}^{-}(|\textbf{r}|) 
- \Phi_{2,\textbf{S}_2}^{+\dagger}(|\textbf{r}|)  \nabla_{\textbf{R}} \Phi_{1,\textbf{S}_1}^{-}(|\textbf{r}-\textbf{R}|)  \biggr)\,,
\end{align}
and the directional derivative, when acting on the spinors, becomes $\nabla_{\textbf{R}} = -\partial/\partial R = -\partial_R $ and can be taken outside the integral.
Subsequently, the integrand  becomes the sum of the eigenspinor products. Following the definition of the eigenspinors in Eq.~(4) of the main text, the general expression of the eigenspinor product is given by 
\begin{align}
\Phi^{+\dagger}_{i,S_i}(r) \Phi^{-}_{j,S_j}(r')
=&   \bra{\uparrow} U^{\dagger}(\textbf{S}_i) (i\sigma_y K) U(\textbf{S}_j)\ket{\uparrow}
   \frac{e^{-(r|\sin{2\delta_i}| + r'|\sin{2\delta_j}|)/\xi}}{ \mathcal{N}_i \mathcal{N}_j k_F^2 r r'} \nonumber \\
&\times\biggr( -\sin{(\delta_i - \delta_j)}\sin{k_F(r+r')} + \sin{(\delta_i + \delta_j)} \sin{k_F(r-r')} \biggr) \,,
\end{align}
where $r,r'$ are scalars.
By substituting Eq.~(S5) into the integrand, the integral becomes 
\begin{align}
I(k_F R, \frac{R}{\xi}) 
& = \partial_R \int d^3 \textbf{r} ~\frac{\bra{\uparrow} U^{\dagger}(\textbf{S}_2) (i\sigma_y K) U(\textbf{S}_1)\ket{\uparrow} - \bra{\uparrow} U^{\dagger}(\textbf{S}_1) (i\sigma_y K) U(\textbf{S}_2)\ket{\uparrow}}{\mathcal{N}_1 \mathcal{N}_2 k_{F}^{2}r_1 r_2} \nonumber \\
&\quad \times \bigg\{ \big(e^{-(r_2|\sin2\delta_{1}|+r_1|\sin2\delta_{2}|)/\xi}+e^{-(r_1|\sin2\delta_{1}|+r_2|\sin2\delta_{2}|)/\xi} \big) \sin(\delta_{1}-\delta_{2})\sin k_{F}(r_1+r_2) \nonumber \\
&\quad \quad \quad  + \big( e^{-(r_2|\sin2\delta_{1}|+r_1|\sin2\delta_{2}|)/\xi}-e^{-(r_1|\sin2\delta_{1}|+r_2|\sin2\delta_{2}|)/\xi} \big)\sin(\delta_{1}+\delta_{2})\sin k_{F}(r_1-r_2) \bigg\} \,,
\end{align}
where $r_1 = |\textbf{r}|, ~r_2 = |\textbf{r}- \textbf{R}|$ can be interpreted as a relative distance from impurity 1 and 2 respectively when the coordinate origin is located at impurity 1. 
The matrix element in the first line only depends on the relative orientation between the two impurities, and can be easily obtained by assuming that the one of the impurities points along the z-axis. 
For the squared matrix element We find   $|\bra{\uparrow} U^{\dagger}(\textbf{S}_2) (i\sigma_y K) U(\textbf{S}_1)\ket{\uparrow} - \bra{\uparrow} U^{\dagger}(\textbf{S}_1) (i\sigma_y K) U(\textbf{S}_2)\ket{\uparrow}|^2 = 2(1-\hat{\textbf{S}}_1 \cdot \hat{\textbf{S}}_2)  =2 F(\textbf{S}_1,\textbf{S}_2)$.

\subsection{Coordinate transformation and integration}

Due to the rotational symmetry along $\hat{\textbf{R}}$, we start by setting up the integral in the cylindrical coordinate $\textbf{r} = (\rho,\theta,z)$. By choosing both impurities to lie on the z-axis and impurity 1 to be at the origin, the relative distance from the two impurities in the cylindrical coordinate are
\begin{align}
\textbf{R} = R \hat{z}, \quad
r_1 = |\textbf{r}| = | \rho \hat{\rho} + z \hat{z}|, \quad
r_2 = |\textbf{r} - \textbf{R}| = |\rho \hat{\rho} + (z - R) \hat{z}|\,.
\end{align}
The measure in the cylindrical coordinate is given by $\int d^3 \textbf{r} = 2 \pi \int_{0}^{\infty} \rho d \rho \int_{-\infty}^{\infty} dz$ which is effectively two dimensional due to rotational symmetry around $\hat{\textbf{R}}$. We then transform  into  elliptical coordinates: $(\rho,z) \rightarrow (r_+,r_-)$ where 
\begin{align}
r_+ = \frac{1}{2}(r_1 + r_2) \, , \quad 
r_- = (r_1 - r_2) \,,
\end{align} 
and the Jacobian is 
\begin{align}
\mathcal{J} = 
\begin{vmatrix}
\partial_{r_+} \rho & \partial_{r_-} \rho  \\ 
\partial_{r_+} z & \partial_{r_-} z  \\ 
\end{vmatrix} 
=
\begin{vmatrix}
\partial_{r_1} \rho & \partial_{r_2} \rho  \\ 
\partial_{r_1} z & \partial_{r_2} z  \\ 
\end{vmatrix}
\begin{vmatrix}
\partial_{r_+} r_1 & \partial_{r_-} r_1  \\ 
\partial_{r_+} r_2 & \partial_{r_-} r_2  \\ 
\end{vmatrix}
= \frac{r_1 r_2}{4 \rho R}\,.
\end{align}
In the second equality we decomposed the coordinate transformation into two steps, i.e. $(\rho,z) \rightarrow (r_1,r_2) \rightarrow (r_+, r_-)$ which results in a product of Jacobians. Note that the Jocobian $\mathcal{J}$ nicely cancels out many factors of the integrand: namely, $\rho$ in the original measure of the cylindrical coordinate, and $r_{1}r_{2}$  that arises from the product of eigenspinors. Expressing Eq.~(S6) in the elliptical coordinates, the integral contains two parts corresponding to the first and the second line inside the parentheses of Eq.~(S6):
\begin{align}
I(k_F R,\frac{R}{\xi})
&= I_1(k_F R,\frac{R}{\xi}) + I_2(k_F R,\frac{R}{\xi}) \,,
\end{align}
where the integrals are straightforward to evaluate, 
\begin{align}
I_1(k_F R,\frac{R}{\xi}) 
&= \partial_R ~ 2\mathcal{Z}\sin{(\delta_1 -\delta_2)}\int_{1/2}^\infty d \rho_+ \int_{-1}^{1}d\rho_-  e^{-2\rho_+ \gamma_+}\cosh{(\rho_-\gamma_-)}\sin{(2R_0\rho_+)}  \\
&= \partial_R ~ 4\mathcal{Z}\sin{(\delta_1 -\delta_2)}e^{-2\gamma_+} \biggl( \frac{ \gamma_+\sin{(2R_0)}+R_0\cos{(2R_0)}}{(\gamma_+^2+R_0^2)} \biggr) \frac{\sinh{(2\gamma_-)}}{2\gamma_-}\,, \\
I_2(k_F R,\frac{R}{\xi}) 
&= \partial_R ~ 2 \mathcal{Z} \sin{(\delta_1 + \delta_2)}\int_{1/2}^\infty d \rho_+ \int_{-1}^{1}d\rho_- e^{-2\rho_+\gamma_+}\sinh{(\rho_-\gamma_-)} \sin{(R_0\rho_-)} \\
&= \partial_R ~ \mathcal{Z} \sin{(\delta_1 + \delta_2)}\frac{e^{-2\gamma_+}}{\gamma_+} (\frac{-2R_0\sinh{(2\gamma_-)} \cos{(2R_0)}+2\gamma_-\cosh{(2\gamma_-)} \sin{(2R_0)}}{R_0^2+\gamma_-^2})\,.
\end{align}
Here we defined  
\begin{gather}
\rho_{\pm} = r_{\pm}/R, \quad R_0= k_F R\,, \quad  \xi_0 = \xi/R  \nonumber \\
\gamma_- = \frac{(|\sin{2\delta_1}|-|\sin{2\delta_2}|)}{2\xi_0}, \quad \gamma_+ = \frac{(|\sin{2\delta_1}|+|\sin{2\delta_2}|)}{2\xi_0}\,, \quad 
\mathcal{Z}=-\frac{\sqrt[]{2F(\textbf{S}_1,\textbf{S}_2)} \pi R}{ \sqrt[]{\mathcal{N}_1 \mathcal{N}_2} k_F^2}\, , \quad \mathcal{N}_i = \frac{(1+\alpha_i^2)}{2\pi \nu \alpha_i \Delta}\,.
\end{gather}

We can next approximate the matrix element by taking into account the typical magnitudes of parameters in real systems. 
For most conventional superconductors with dilute magnetic impurities, the Fermi wavelength is much smaller than the superconductor coherence length and impurity separation. 
We can thus expand $I(k_F R,R/\xi)$ to lowest order in $1/(k_F \xi)$ and  $1/(k_F R)$.  
The approximate matrix element is then given by 
\begin{align}
 | \mathcal{M}^{1,2}_{\text{YSR}} | 
&\approx  \Delta ~ \sqrt[]{2F(\textbf{S}_1,\textbf{S}_2)} \frac{e \vec{\mathcal{E}} \cdot \hat{\textbf{R}} \xi}{\hbar \omega} \frac{1}{  k_F R} \, \sqrt[]{\frac{\alpha_1}{1+\alpha_1^2} \frac{\alpha_2}{1+\alpha_2^2}}  \sinh \bigg\{ \frac{R(|\sin 2\delta_1|-|\sin 2\delta_2|)}{\xi} \bigg\}  \sin(2k_F R)  \nonumber \\
& \quad \times e^{-(|\sin(\delta_1)|+|\sin(\delta_2)|)\frac{R}{\xi}}\bigg( \frac{\sin(\delta_1-\delta_2)}{|\sin 2\delta_1|-|\sin 2\delta_2|} -\frac{\sin(\delta_1+ \delta_2)}{|\sin 2\delta_1|+|\sin 2\delta_2|}  \bigg) \,.
\end{align}
Note that in the last line, we only keep the leading term of many terms generated by the $R$-derivative because $k_F \gg 1/\xi$: namely, $\partial_R \cos(2k_F R)$. 

\subsection{The narrow band approximation}
For an ensemble of moments that forms a narrow YSR band, the exchange coupling of any two impurities in the ensemble are very close to each other which can be quantified precisely as 
$\big| |\sin 2\delta_i|-|\sin 2\delta_j| \big|\ll \xi/R$ for any YSR pair $(i,j)$ in the ensemble separated by distance $R$. Without loss of generality, we assume that $\alpha \in [0,1]$, so $\delta \in [0,\pi/2]$ and $\sin{2\delta } > 0$.
Using the following trigonometric identities and  approximation 
\begin{align}
\sin 2\delta = \frac{2\alpha}{1+\alpha^2},~ \cos2\delta = \frac{1-\alpha^2}{1+\alpha^2} ,  \quad \sin(2\delta_2)-\sin(2\delta_1) \approx \frac{d \sin(2\delta)}{d \alpha} \bigg|_{\alpha = \bar{\alpha}} (\alpha_2-\alpha_1)   \,,
\end{align}
we can expand the matrix element to the first order in ($\alpha_2 - \alpha_1$) and retrieve Eq.~(8) of the main text.

\section{The integrated conductivity}

\maketitle
From Eqs.~(8)-(9) of the main text, the normalized dissipative part of the conductivity for an impurity ensemble is 
\begin{equation}
\frac{\sigma(\omega)}{\sigma_{n}}
\approx   \frac{1}{\mathcal{V}\hbar\omega}\,\frac{ 1 }{\hbar \pi \nu^2 \xi^2 l  \Delta^2}  \frac{\alpha^4}{(1+\alpha^2)^2} \sum_{i \neq j}F(\textbf{S}_i \cdot \textbf{S}_j) \cos^2\theta_{ij} \, (E_i - E_j)^2  \sin^2(k_F R_{ij}) \, e^{-\frac{4\alpha}{1+\alpha^2}\frac{R_{ij}}{\xi}}\,\delta(E_{i}+E_{j}-\hbar\omega) \,.
\end{equation}

We introduced $\cos \theta_{ij} = \hat{\mathcal{E}} \cdot \hat{\textbf{R}}_{ij} $  the cosine of the mutual angle between the applied electric field $\vec{\mathcal{E}}$ and $\textbf{R}_{ij} = \textbf{R}_i - \textbf{R}_j$. 
This factor ensures that only the field component parallel to $\textbf{R}_{ij}$ contributes to absorption. 
We can rewrite the summation as   
\begin{align}
&\sum_{i \neq j}F(\textbf{S}_i \cdot \textbf{S}_j)\cos^2\theta_{ij} \, (\alpha_i - \alpha_j)^2  \sin^2(k_F R_{ij}) \, e^{-\frac{4\alpha}{1+\alpha^2}\frac{R_{ij}}{\xi}}\,\delta(E_{i}+E_{j}-\hbar\omega) 
\nonumber \\
&=\int d\varepsilon d\varepsilon'  (\varepsilon - \varepsilon')^2  \delta(\varepsilon+\varepsilon'-\hbar\omega) \sum_{i \neq j} \delta(\varepsilon_{i}-\varepsilon) \delta(\varepsilon_{j}-\varepsilon') F(\textbf{S}_i \cdot \textbf{S}_j) K(|\textbf{R}_i-\textbf{R}_j|)\,,
\end{align}
where we  introduced 
\begin{align}
    K(|\textbf{R}_i-\textbf{R}_j|) =
     e^{-4|\sin2\delta||\textbf{R}_i-\textbf{R}_j|/\xi}\sin^{2}(2k_{F}|\textbf{R}_i-\textbf{R}_j|) \,  \cos^2 \theta_{ij}\,. \label{eq:sigma2}
\end{align}

Next we average $\sigma(\omega)$  over the  distribution of Shiba energies, 
$\langle \sigma(\omega) / \sigma_{n} \rangle = \prod_i [\int d \varepsilon_i \rho(\varepsilon_i)] \frac{\sigma(\omega)}{\sigma_{n}}$. Here we consider consider a  uniform density of states $\rho(\varepsilon)=W^{-1}\Theta(\frac{1}{2}W-|\varepsilon-E_{\text{YSR}}|)$ in a band of width  $W$ centered at $E_{\text{YSR}}$. 
Upon averaging in Eq.~(\ref{eq:sigma2}), the integral over energies and the sum over positions decouple. 
The energy integral becomes
\begin{align}
 \int d\varepsilon d\varepsilon' (\varepsilon-\varepsilon')^2\rho(\varepsilon)\rho(\varepsilon') \delta(\varepsilon+\varepsilon'-\hbar\omega)
&=  \frac{2 }{3W^2} (W -|\hbar \omega - 2\bar{E}_{\text{YSR}} |)^3 \Theta(W - |\hbar \omega - 2\bar{E}_{\text{YSR}} |)\,,
\end{align} 
while the sum over positions is 
\begin{equation}
   \langle \sum_{i \neq j}   F(\textbf{S}_i \cdot \textbf{S}_j) K(|\textbf{R}_i-\textbf{R}_j|) \rangle = N \bar{F}(\{ {\textbf{S}(\textbf{r}}) \}) \,, \quad N= \langle \sum_{i \neq j}   K(|\textbf{R}_i-\textbf{R}_j|) \rangle\,,
\end{equation}
which defines $\bar{F}(\{ {\textbf{S}(\textbf{r}}) \})$ of the main text. 
The total conductivity becomes 
\begin{equation}
\bigg\langle \frac{\sigma(\omega)}{\sigma_{n}} \bigg\rangle = \frac{\bar{F}(\{ {\textbf{S}(\textbf{r}}) \})}{\hbar\omega}\,\frac{2 \alpha^4}{3 (1+\alpha^2)^2} \frac{n^2}{\hbar \pi \nu^2 \Delta} \frac{\xi^{D-2}w^{3-D}}{l}  \frac{1}{W^2 \Delta} (W -|\hbar \omega - 2\bar{E}_{\text{YSR}} |)^3  \Theta(W - |\hbar \omega - 2\bar{E}_{\text{YSR}} |)  \,.
\end{equation}
If we integrate this ratio over all frequency below the superconducting gap and expand the integral to the lowest order in $W/E_{\text{YSR}}$ (narrow-band approximation), we find 
\begin{align}
\int_0^{2\Delta} \bar{F}(\{ {\textbf{S}(\textbf{r}}) \})\bigg\langle \frac{\sigma(\omega)}{\sigma_{n}} \bigg\rangle  d (\hbar \omega) 
&\approx \bar{F}(\{ {\textbf{S}(\textbf{r}}) \})\, \frac{n^2}{\hbar \nu^2 \Delta} \frac{\xi}{l}  \bigg(\frac{w}{\xi} \bigg)^{3-D}  \bigg(\frac{W}{\Delta} \bigg)^2 \frac{2\alpha^4}{12 \pi (1-\alpha^4)}  \,,
\end{align}
which gives us the equivalent of Eq.~(1) of the main text in the clean, dilute limit, (see also Fig.~2 of the main text).

\section{Radiation-induced exchange vs RKKY interaction}
In this section, we will calculate for the induced  YSR interaction the mean $\langle J_{\text{YSR}}^{\text{ind}} \rangle$ and the variance $\delta J_{\text{YSR}}^{\text{ind}}$.
We also estimate the ratio of $\langle J_{\text{YSR}}^{\text{ind}} \rangle$ to the typical RKKY exchange coupling. 

From Eq.~(11) of the main text, the effective spin-spin coupling obtained from the AC stark shift is given by  
\begin{align}
J_{\text{YSR}}^{\text{ind}} (R)
&= B(\alpha,\mathcal{E},\omega) (\alpha_2 - \alpha_1)^2 \,, \\
B(\alpha,\mathcal{E},\omega) 
&= \frac{4}{3}\frac{\Delta^2}{2E_{\text{YSR}}-\hbar \omega} \bigg( \frac{e \, \mathcal{E}  \xi  }{\hbar \omega} \bigg)^2  \frac{1}{(k_F \xi)^2} \bigg( \frac{\alpha}{1+\alpha^2} \bigg)^6   e^{-\frac{4\alpha}{1+\alpha^2}\frac{R}{\xi}} \,.
\end{align}
Here we averaged the coupling strength over  orientations, $\cos^2 \theta_{12}  \approx 1/3$, and short distances, $\sin^2(k_F R) \approx 1/2$.

Next, we use $\alpha_2 - \alpha_1 = \big[ d \alpha/dE_{\text{YSR}} \big|_{\alpha=\bar{\alpha}} \big] (\epsilon_2 -\epsilon_1)$,  assuming a narrow band and  approximating $d \alpha/dE_{\text{YSR}} \big|_{\alpha=\bar{\alpha}} = -(1+\bar{\alpha}^2)^2/4\bar{\alpha}\Delta$ by a constant.
By averaging   the energies $\epsilon_{1,2}$ over a narrow uniform band, we find the mean and the variance, 
\begin{align}
\langle J_{\text{YSR}}^{\text{ind}} \rangle
&\approx - B(\alpha,\mathcal{E},\omega) 
 \bigg( \frac{d \alpha}{dE_{\text{YSR}}} \bigg)^2 \frac{W^2}{6}
= -\bigg\{ \frac{ 1}{2E_{\text{YSR}}-\hbar \omega} \bigg( \frac{e \, \mathcal{E}  \xi  }{\hbar \omega} \bigg)^2  \frac{1}{(k_F \xi)^2}  \frac{\alpha^4}{(1+\alpha^2)^2}~  e^{-\frac{4\alpha}{1+\alpha^2}\frac{R}{\xi}} \bigg\}  \frac{W^2}{12}\,,
\\
\langle (J_{\text{YSR}}^{\text{ind}})^2 \rangle
&\approx  B(\alpha,\mathcal{E},\omega)^2 
\bigg( \frac{d \alpha}{dE_{\text{YSR}}} \bigg)^4 \frac{W^4}{15}\,,
\end{align}
where $\delta J_{\text{YSR}}^{\text{ind}} = \sqrt{ \langle (J_{\text{YSR}}^{\text{ind}})^2 \rangle - \langle J_{\text{YSR}}^{\text{ind}} \rangle^2}$. 
From Eq.~(S28) we see that the fluctuation of $J_{\text{YSR}}$ is of the  order of its mean,
$\delta J_{\text{YSR}}^{\text{ind}}/ |\langle J_{\text{YSR}}^{\text{ind}} \rangle| = \, \sqrt[]{7/5}$. 
This is expected since we saw in the main text that the absorption matrix element vanishes in the reflection symmetric case, and  we broke that symmetry by introducing unequal couplings $\alpha_{1,2}$.  

The expression for RKKY interaction in the weak exchange limit ($\alpha \ll 1$) is obtained from the first part of Eq.~(13) from Ref.~\cite{norman},
\begin{align}
\langle J_{\text{RKKY}} \rangle &= \frac{E_F \alpha^2}{\pi (k_F R)^3} \frac{1}{\sqrt[]{2}} e^{-2R/\xi} F_1[\frac{2R}{\xi}] \,,
\end{align}
where $F_1[\beta] =  \beta \int_0^{\infty} e^{-\beta (\sqrt[]{x^2+1}-1)} dx $ is a numerical factor of order $\mathcal{O}(1)$ and the purely antiferromagnetic correction is omitted since we are interested in the regime $R \lesssim \xi$. Using Eqs.~(S27),~(S29), the ratio of the induced coupling to RKKY coupling in the weak coupling ($\alpha \ll 1$) regime is
\begin{align}
\bigg| \frac{\langle J_\text{YSR}^{\text{ind}} \rangle } {\langle J_\text{RKKY} \rangle } \bigg|
&\leq \frac{\sqrt[]{2} \, \pi}{12} \frac{W}{\Delta} \bigg( \frac{R}{\xi} \bigg)^3  \big( \frac{\alpha}{1+\alpha^2} \big)^2   e^{-2\frac{R}{\xi}(|\sin 2\delta |- 1)}\,,  \quad ( \frac{e \mathcal{E} }{\hbar \omega} = \frac{1}{\xi} \, ,~ 2E_{\text{YSR}}-\hbar \omega = 2W )\,.
\end{align}
In the above expression we obtained the upper bound of the coupling ratio by setting the electric field strength to its maximum value allowed by the critical supercurrent, 
and the detuning to be equal to the YSR bandwidth. 
In practice the detuning needs to be larger than the bandwidth in order to avoid resonant driving.

Figure~S1 compares the narrow-band approximation of the coupling ratio,
Eq.~(S30),
and  the  exact numerical average of the ratio obtained from [the square of] Eq.~(S16) and  Eq.~(S29), without assuming a narrow band. 
Even though the bandwidth is not necessarily small compared to $2 E_{\text{YSR}}$, we find good agreement between the narrow band approximation and the numerical integration even outside the regime of expected validity.
We see  that the induced interaction can be as strong as RKKY interaction for $\bar{\alpha} \approx 0.2 - 0.7$ in the moderately dilute limit $R =3 \xi$. 

\begin{figure}
\centering
\includegraphics[width=15cm]{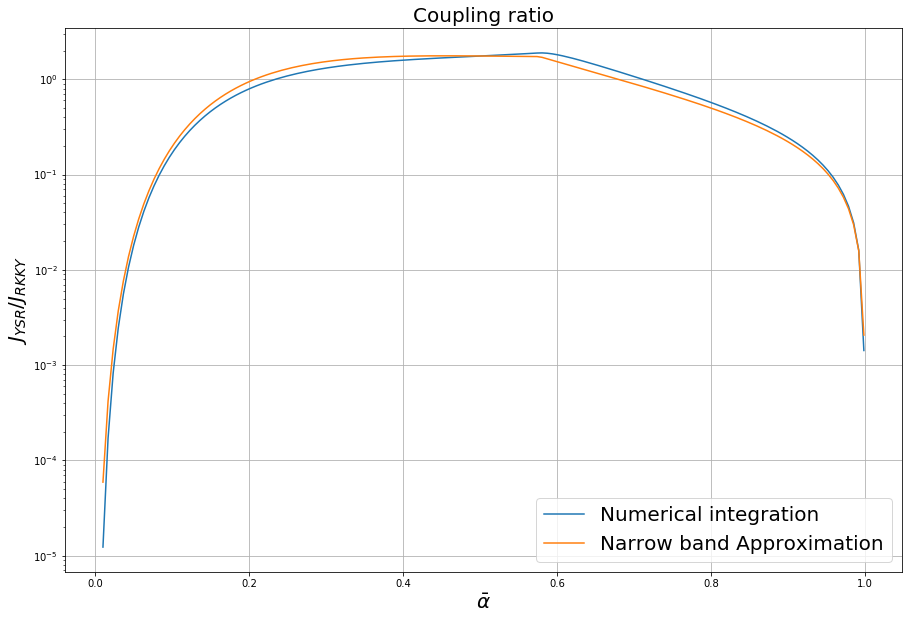}
\caption{
The ratio of the radiation-induced and RKKY exchange couplings as a function of the average dimensionless coupling $\bar{\alpha}$ which sets the position of the Shiba band center. 
Here we consider the dilute clean limit, with the average distance between the magnetic moments $R = 3\,\xi$.
The electric field magnitude and detuning are chose as outlined below Eq.~(S30). 
The bandwidth $W$ is different for each value of $\bar{\alpha}$ and chosen to be the maximum allowed value for a sub-gap level, i.e. $W/2 = \min[\Delta -E_{\text{YSR}}(\bar{\alpha}) , E_{\text{YSR}}(\bar{\alpha})]$. 
The orange curve is obtained from Eq.~(S30) which is derived in the narrow-band approximation, $W\ll E_{\text{YSR}}$.
The  blue curve is obtained from numerical integration using Eq.~(S16) and averaging over the uniform density of states $\rho(E)$ without any assumption about $W/E_{\text{YSR}}$.
We see that in the dilute limit,  the radiation-induced interaction may even exceed to  RKKY interaction strength.
}
\end{figure}